\documentclass[conference,twocolumn,10pt]{IEEEtran}

\usepackage[utf8]{inputenc}
\usepackage{enumerate}
\usepackage{amsmath}
\usepackage{algpseudocode}
\usepackage{algorithm}
\usepackage{amssymb}
\usepackage{color}
\usepackage{graphicx}
\usepackage{amsmath}
\usepackage{subfigure}
\usepackage{multirow}

\usepackage{lipsum}
\usepackage{xcolor}
\usepackage{makecell}
\usepackage{mathtools}
\usepackage{comment}
\usepackage{optidef}
\usepackage{cite}
\usepackage{float}
\DeclareGraphicsExtensions{.pdf,.jpeg,.png,.eps}
\usepackage{caption}
\DeclarePairedDelimiter\ceil{\lceil}{\rceil}

\DeclareUnicodeCharacter{2061}{}
\newcommand\blfootnote[1]{%
  \begingroup
  \renewcommand\thefootnote{}\footnote{#1}%
  \addtocounter{footnote}{-1}%
  \endgroup
}

\begin{document}
\graphicspath{ {./images/} }
\DeclareGraphicsExtensions{.pdf,.png,.jpg,}

\title{Energy-Efficient Multi-UAV Data Collection for IoT Networks with Time Deadlines}


\author{\IEEEauthorblockN{Oussama Ghdiri, Wael Jaafar, Safwan Alfattani, Jihene Ben Abderrazak, and Halim Yanikomeroglu}
}



\maketitle
\begin{abstract}
In this paper, we focus on 
energy-efficient UAV-based IoT data collection in sensor networks in 
which the sensed data have different time deadlines. In the 
investigated setting, the sensors are clustered and managed by cluster 
heads (CHs), and multiple UAVs are used to collect data from the CHs.
The formulated problem is solved through a two-step approach. In the first step, 
\textcolor{black}{an efficient method is proposed to determine the minimal number of CHs and their best locations.}
Subsequently, the minimal number of UAVs and their trajectories are obtained by solving the associated capacitated vehicle routing problem. Results show the efficiency of our proposed 
CHs placement method compared to baseline approaches, where bringing the CHs closer to the dockstation allows significant energy savings.
Moreover, 
\textcolor{black}{among different UAV trajectory planning algorithms, Tabu search achieves the best energy consumption.}
Finally, the impact of the battery capacity and time deadline are investigated in terms of consumed energy, number of visited CHs, and number of deployed UAVs.
\end{abstract}

\blfootnote{O. Ghdiri and J. Ben Abderrazak are with ESPRIT School of Engineering Tunis, Tunisia, W. Jaafar and H. Yanikomeroglu are with Carleton University, Canada, and S. Alfattani is with the University of Ottawa, Canada, and King Abdulaziz University, Saudi Arabia.

The 
authors would like to thank Professor Abbas Yongacoglu, University of 
Ottawa, Canada, for his comments about this work.}

\section{Introduction}
The vision of future wireless technologies with numerous smart applications and services has emerged with  the Internet
of Things (IoT), connecting millions of devices and people. IoT sensor networks are a key enabler for novel applications, including smart cities, connected cars, smart grids, and intelligent transportation. Indeed, the Third Generation Partnership Project (3GPP) has issued recently several standardization documents to support IoT in 5G systems (e.g., TR 23.700-20 and TR 23.700-24). Typically, IoT sensors are low-cost and limited-energy devices with short transmission ranges. In large-scale IoT systems, the deployment of super-sensors or cluster heads (CHs) to aggregate data from neighboring sensors is a practical solution to reduce communication overhead and save sensors' energy. However, even CHs may have a limited communication range, and therefore, a data collector needs to be deployed to gather sensed data and send it to the control center. In that matter, UAVs can be deployed as data collectors \cite{DBLP:conf/globecom/AlfattaniJYY19}. Indeed, UAVs 
have been presented as an efficient tool to solve several wireless communication challenges \cite{Alzenad2017,cherif2020optimal,jaafar2020multiple}.
Accordingly, recent 3GPP standards were issued to regulate different aspects of UAV support in 5G (TR 22.829 and TS 22.125).
Due to the agility, strong Line-of-Sight (LoS), and fast deployment of UAVs, the latter are a relevant option for data collection in IoT systems \cite{DBLP:conf/globecom/AlfattaniJYY19}.
Nevertheless, UAVs have a limited onboard energy and are sensitive to blockages. Hence, accurate communication channel and power consumption models should be considered when designing data collection missions. On the other hand, large-scale IoT networks encompass sensors with different priorities, criticality levels, and time-sensitive usefulness. Consequently, energy-efficient data collection needs an accurate UAV deployment and trajectory planning to meet time-sensitive constraints, e.g., time deadlines for data collection. 

Several works proposed UAV-based sensors' data collection. In our previous work \cite{DBLP:conf/globecom/AlfattaniJYY19}, we proposed collecting data from a clustered wireless sensor network using one or several UAVs, aiming to minimize the mission costs in terms of number of clusters and traveled distance by UAVs, while guaranteeing data is collected from all clusters. In this setting, sensor nodes (SNs) were clustered using a K-means-based algorithm, and UAV trajectories designed with different approaches, including genetic algorithm and nearest-neighbor. In \cite{DBLP:journals/twc/SamirSANG20}, the problem of minimizing data collecting time from IoT time-constrained sensors is investigated, where the trajectory of a single collecting UAV and radio resource allocation are optimized.  
Moreover, the authors of \cite{DBLP:journals/wcl/ZhanZZ18} proposed joint optimization of wakeup schedule of SNs and a single UAV's trajectory to minimize the total consumed energy by SNs, while guaranteeing complete data collection from all of them.  
Finally, authors in \cite{Salih2018} presented a data collection framework, where multiple UAVs are deployed to collect data from clustered SNs, aiming to minimize the number of deployed UAVs and their flight times and distances, with respect to a mission deadline. The optimal solution is obtained, which outperforms a benchmark greedy approach. 
Nevertheless, these works have several limitations. Authors in \cite{DBLP:conf/globecom/AlfattaniJYY19,DBLP:journals/twc/SamirSANG20} did not consider energy restrictions of UAVs, while \cite{DBLP:journals/wcl/ZhanZZ18} focused only on single UAV trajectory planning. Finally, \cite{Salih2018} did not account for energy limitations and air-to-ground accurate channel modeling.

Motivated by the aforementioned points, we propose here a practical data collection approach for large-scale IoT systems, where multiple UAVs collect sensed data with different time deadlines.  
The proposed solution follows a two-step approach. First, the minimum number of CHs and their locations are determined to guarantee that data from all SNs is aggregated at the CHs. Then, the number of collecting UAVs and their trajectories are optimized to minimize the total consumed energy. 
Finally, the impact of key parameters, e.g., UAV's battery capacity and data time deadlines, is investigated. 

\section{System Model}

\subsection{Network and Channel Models}
We consider an IoT sensor network consisting of a set $\mathcal{M} = \{1,\ldots,M\}$ of SNs, randomly located in a large area and constantly collecting time-sensitive data.
We assume that SNs are heterogeneous, i.e., sensed data by SNs are of different kind, having different importance in time. Indeed, some sensed data are more critical and need to be collected faster than other less-critical data.
Due to the limited energy and communication capability of typical IoT sensors, we assume that a set of $\mathcal{K} = \{1,\ldots,K\}$ cluster heads (CHs) can be deployed to collect, aggregate, and transmit sensed data to the data collector. In our system, collection of data is operated by multiple UAVs, which fly and hover above/close to CHs for a sufficient time to collect the aggregated data.

Each sensor transmits its sensed data to its associated CH using power $P_{\rm SN}$, while a CH communicates with the collecting UAV using power $P_{\rm CH}>P_{\rm SN}$. Assuming that IoT sensors communicate with their associated CH using the time-division multiple access (TDMA) protocol,
the received signal-to-noise ratio (SNR) at CH $c$ for the signal transmitted by SN $s$, denoted $\gamma_{sc}$, can be written as
\begin{equation}
\small
    \label{eq:SNR_chi}
  \gamma_{sc}=\frac{P_{\rm SN} d_{sc}^{-\alpha}}{\sigma^2}   ,\forall s \in \mathcal{M}, \; \forall c \in \mathcal{K}  
\end{equation}
where $d_{sc}=||\textbf{q}_s-\textbf{q}_c||$ is the distance between IoT sensor $s$ and CH $c$, $\textbf{q}_i=[x_i,y_i,z_i]$ is the 3D location of node $i$ ($i=s$ or $c$), $\alpha$ is the path loss exponent, and $\sigma^2$ is the noise power.
This communication link is considered successful if $\gamma_{sc}\geq \gamma_{\rm{th}}$, where $\gamma_{\rm{th}}$ is a selected SNR threshold. Hence, a maximum communication range for each SN $s$ associated with CH $c$ is defined as \cite{DBLP:conf/globecom/AlfattaniJYY19}
\begin{equation}
    \label{eq:distance_SN_CH}
    \small
    d_{sc}\leq d_{\rm{th}}=\left(\frac{P_{\rm SN}}{\sigma^2 \gamma_{\rm{th}} }\right)^{1/\alpha}  ,\forall s \in \mathcal{A}_c,\; c \in \mathcal{K}
\end{equation}
where $\mathcal{A}_c$ is the set of SNs associated with CH $c$.

Given (\ref{eq:distance_SN_CH}), CHs are deployed to collect data from SNs and send it to UAVs.
Let $\mathcal{U}=\{1,\ldots,U\}$ be the set of available UAVs for CHs data collection. 
For the sake of simplicity, we assume that a UAV flies at a fixed altitude $H$ above ground, such that $H$ respects authority regulations and safety considerations. 
Also, UAVs are equipped with sensors that allow obstacle avoidance at a certain safety distance, denoted $d_{\rm safe}$, when they fly at maximum speed $v_{\max}$.

When UAV $u$ hovers to receive data from CH $c$, the air-to-ground channel between them can be expressed using the probabilistic path loss model given by \cite{DBLP:journals/wcl/Al-HouraniSL14}
\begin{equation}
\small
\label{eq:channel_path_loss}
    \Lambda_{cu}=Pr_{cu}^{\rm{LoS}} l_{cu}^{\rm{LoS}}+ Pr_{cu}^{\rm{NLoS}} l_{cu}^{\rm{NLoS}}, \; \forall c \in \mathcal{K}_u, \; \forall u \in \mathcal{U},
\end{equation}
where $\Lambda_{cu}$ is the average path-loss between CH $c$ and UAV $u$, $\mathcal{K}_u \subset \mathcal{K}$ is the subset of ordered CHs to visit by UAV $u$, $Pr_{cu}^{\rm{LoS}}$ is the LoS probability, $Pr_{cu}^{\rm{NLoS}}=1-Pr_{cu}^{\rm{LoS}}$ is the NLoS probability, and $l_{cu}^{\rm{LoS}}$ and  $l_{cu}^{\rm{NLoS}}$ are LoS and NLoS path-loss, respectively. These parameters are written as \cite{DBLP:journals/wcl/Al-HouraniSL14}
\begin{equation}
\small
\label{eq:prob_path_loss}
    Pr_{cu}^{\rm{LoS}}={1}/{\left(1+a e^{b \ \left(\theta_{cu}-a\right)}\right)},\; \forall c \in \mathcal{K}_u, \; \forall u \in \mathcal{U},
\end{equation}
and
\begin{equation}
\small
\label{eq:path_loss}
    l_{cu}^m=20 \log(4\pi f_c / V)+20 \log(d_{iu})+\beta_m,\; m =\rm{LoS} \text{ or } \rm{NLoS}
\end{equation}
where $a$ and $b$ are constant values determined from the environment (rural, urban, etc.), $\theta_{cu}=\frac{180}{\pi} \times \arcsin \left(\frac{H}{d_{cu}}\right)$ is the elevation angle between UAV $u$ and CH $c$, $f_c$ is the carrier frequency, $V$ is the speed of light, and $\beta_m$ is the excessive path-loss coefficient.
Using Friis formula, the received power at UAV $u$ from CH $c$ is expressed by
\begin{equation}
\small
\label{eq:power_uav_ch}
    P_{cu}=P_{\rm CH}-\Lambda_{cu} 
    ,\; \forall c  \in \mathcal{K}_u,\; \forall u   \in \mathcal{U},
\end{equation}
\textcolor{black}{where $P_{\rm CH}$ is the unit power (in Watt), i.e. equal to 0 dBm.}

\subsection{UAV Data Collection Model}



We assume that the mission time, i.e., the maximum time for UAVs to collect data from CHs and return to their dockstation, is $T_F$ and divided into $N$ time slots (TSs). Thus, TSs are equal and are of length $\delta=\frac{T_F}{N}$. 
We define the 3D location of UAV $u$ in time slot $t$ as $\textbf{q}_u^t=[x_u^t, y_u^t, H]$. Since the UAV's speed is limited by $v_{\max}$, its traveled distance in one TS is constrained by
\begin{equation}
\small
\label{eq:horizontal_distance}
    ||\mathbf{q}_u^{t+1}-\mathbf{q}_u^{t}||=\sqrt{(x_u^{t+1}-x_u^t)^2+(y_u^{t+1}-y_u^t)^2} \leq v_{\rm{max}} \delta.
\end{equation}


In our system, each CH $c$ uploads its data to its associated UAV $u$ at rate $R_{cu}$ in bits/second (bps), determined using the Shannon equation as follows:
\begin{equation}
\small
\label{eq:transmission_rate}
    R_{cu}= W \log_2(1+\gamma_{cu}), \; \forall c \in \mathcal{K}_u, \; \forall u \in \mathcal{U},
\end{equation}
where $W$ is the total bandwidth of the channel and $\gamma_{cu}=\frac{\bar{P}_{cu}}{\sigma^2}$ is the SNR, \textcolor{black}{where $\bar{P}_{cu}=10^{\frac{P_{cu}}{10}}$ is the linear value of $P_{cu}$}. Accordingly, the required number of TSs to upload one packet to the UAV can be expressed by
\begin{equation}
\small
\label{eq:transmission_time_4_1packet}
T_{1,cu}= \ceil*{\frac{S_{p}}{R_{cu}\delta}}, \; \forall c \in \mathcal{K}_u, \; \forall u \in \mathcal{U},
\end{equation}
where $S_{p}$ is the size of the packet in bits and $\ceil{\cdot}$ is the ceiling function. Given $Q_c$ data packets to be transmitted by CH $c$, then the required data collection time (in TSs) by UAV $c$ is
%
\begin{equation}
\small
\label{eq:collectionTime1UAV}
    T_{cu}={Q_c}{T_{1,cu}}\; \forall c \in \mathcal{K}_u, \; \forall u \in \mathcal{U}.
\end{equation}

Due to the different priorities of collected data 
at each CH, we define by $T_{d,c}$ the time deadline of data stored in CH $c$, i.e., if not totally collected within this deadline, the data present in the CH becomes outdated and is dropped\footnote{Typically, this deadline is determined by the most critical sensed data.}. Hence, 
data collection has to respect the following constraint, assuming that all packets in one CH are collected by only one UAV
\begin{equation}
\small
\label{eq:deadline1}
    \sum_{\substack{m=1}}^{\phi_u(c)} \left[ T_{mu}+ T_{f,u}(m-1,m) \right]\leq T_{d,c},\; \forall c \in \mathcal{K}_u,  
\end{equation}
where $\phi_u(c)$ designates the rank (i.e. index) of CH $c$ in $\mathcal{K}_u$, $T_{mu}$ is the data collection time of CH ranked $m$ in $\mathcal{K}_u$, and $T_{f,u}(m-1,m)$ is the flight time (in TSs) from the hovering location to collect data from CH ranked $(m-1)$ to that associated to the CH ranked $m$ in $\mathcal{K}_u$, with $m=0$ designating the dockstation. The flight time of UAV $u$ between two locations $\textbf{q}_{u,i}$ and $\textbf{q}_{u,i'}$, where $i$ and $i'$ are two CHs ranks in the UAV path, can be calculated as follows:
\begin{equation}
\small
\label{eq:flightTime}
T_{f,u}(i,i')=\frac{||\mathbf{q}_{u,i'}-\mathbf{q}_{u,i}||}{v_u\; \delta}, 
\end{equation}
with $v_u \leq v_{\max}$ is the flying speed of the UAV (in m/sec).


\subsection{UAV Energy Model}

The UAV is energy-constrained due to limited on-board battery. 
The battery lifetime depends on several factors, e.g.,  UAV's energy source, type, weight, speed, etc.
Typically, the UAV’s energy consumption consists of the propulsion energy and the communication energy. 
Without loss of generality, communication energy is several orders of magnitude smaller than propulsion energy. Hence, it is neglected in the considered energy model.
For the propulsion energy, we adopt the propulsion-power model for rotary-wing UAVs of \cite{DBLP:journals/twc/ZengXZ19}
\begin{equation}
\small
\label{eq:propulsion_power}
\begin{split}
P_{\rm{prop},u}(v_u) = & \;\zeta_I \left(\sqrt{1+\frac{v_u^4}{4v_0^4}}-\frac{v_u^2}{2v_0^2}\right)^{1/2}  \\ 
+ & \; \zeta_B \left( 1+\frac{3v_u^2}{U_{\rm tip}^2} \right)  + \frac{1}{2} d_0 \psi r A v_u^3,
\end{split}
\end{equation}
where $\zeta_B$ and $\zeta_I$ are the blade profile power and induced power, respectively. $U_{\rm tip}$ is the tip speed of the rotor blade, $v_0$ is the mean rotor induced velocity, $\psi$ is the air density, $d_0$ is the fuselage drag ratio, $r$ is the rotor solidity, and $A$ denotes the rotor disc area. 
In order to obtain the power consumption for hovering, we use the same equation above, but with null flying speed $v_u = 0$, thus, we get
\begin{equation}
\small
\label{eq:hovering_power}
P_{\rm{hov},u}=P_{\rm{prop},u} (v_u=0)=\zeta_B+\zeta_I.
\end{equation}
Using (\ref{eq:propulsion_power})--(\ref{eq:hovering_power}), the consumed energy of UAV $u$ during a time period $T$ can be given by
\begin{equation}
\small
\label{eq:energy}
E_{u}(v,T) =  
  \begin{cases}
    P_{\rm{prop},u}(v) \times T, & \text{ if } v>0 \\
    P_{\rm{hov},u} \times T,  & \text{ if } v=0.\\
  \end{cases}
\end{equation}
Hence, the related battery status at TS $t$, denoted $S_u(t)$, can be expressed by
\begin{equation}
\small
    S_u(t)=S_u(t-1)-E_u(v,\delta), \; \forall t> 1,
\end{equation}
where $S_u(t-1)$ is the battery's status at the end of TS $(t-1)$, and $S_u(0)$ is the initial battery capacity. The latter is expressed as $S_u(0)=S_{\rm{ini}}+S_{\min}$, where $S_{\rm{ini}}$ is the battery capacity dedicated for the mission, while $S_{\min}$ is a safety capacity, reserved for emergency pull back to the dockstation. Hence, $S_u(t) \in [S_{\min}, S_u(0)]$.


\section{Problem Formulation}

In this section, we formulate our optimization problem aiming to minimize the total consumed energy for data collection, by optimizing the deployment of CHs, the number of required UAVs, and their trajectories. Let $K_u=|\mathcal{K}_u|$ be the cardinality of the set $\mathcal{K}_u$, $\forall u \in \mathcal{U}$. Then, using (\ref{eq:energy}), the consumed energy by UAV $u$ during the mission is given by
\begin{equation}
\small
    \label{eq:consumedUAV}
    E_u(\mathcal{K}_u)=\sum_{m=1}^{K_u+1} \left[E_u(0,T_{mu} \; \delta) + E_u(v_u,T_{f,u}(m-1,m) \; \delta)  \right],
\end{equation}
where $T_{f,u}(K_u,K_u+1)={\|\textbf{q}_{u,K_u}-\textbf{q}^0\|}/{v_u \delta}$, and $\textbf{q}^0$ is the initial location of all UAVs, i.e., the dockstation.
Therefore, the optimization problem can be formulated as follows:

\begin{subequations}
    \small{\begin{align}
    &
    \label{p1}   \min_{ \substack{\mathcal{K},\mathcal{L},\{ \mathcal{A}_c\}_{c \in \mathcal{K}}\\ \mathcal{U}, \left\{\mathcal{K}_u,\mathcal{L}_u\right\}_{u \in \mathcal{U}} }}  \quad 
	\sum_{u=1}^{U} E_u(\mathcal{K}_u) \tag{P1} \\
	\text{s.t.} \quad
	\label{c2} & d_{sc} \leq d_{\rm{th}}, \; \forall s \in \mathcal{A}_c , \forall c \in \mathcal{K},  \tag{P1.a}\\
	\label{c1}  & S_u(t) \geq S_{\min}, \; \forall u \in \mathcal{U}, \forall 1 \leq t \leq T_u, \; \forall u \in \mathcal{U} \tag{P1.b} \\
	\label{c3} &  \sum_{m=0}^{\phi_u(c)} \left( T_{mu}+ T_{f,u}(m-1,m)\right) \leq T_{d,c}, \tag{P1.c}\\ 
	\label{c4} & \| \mathbf{q}_u^{t}-\mathbf{q}_{u'}^t \| \geq d_{\rm{safe}},  \forall t \geq 1, \forall (u,u') \in \mathcal{U}^2,\; u \neq u' \tag{P1.d}\\
	\label{c5} & \| \mathbf{q}_u^{t+1}-\mathbf{q}_u^t \| \leq v_{max} \delta,  \; \forall 1 \leq t \leq T_u, \; \forall u \in \mathcal{U} \tag{P1.e},
    \end{align}}
\end{subequations}

\noindent
where $T_u \leq N$ is the effective mission completion time for UAV $u$ (in TSs), $\mathcal{L}=\{ \textbf{q}_c\}_{c \in \mathcal{K}}$ is the set of selected locations for CHs, and $\mathcal{L}_u=\{ \textbf{q}_{u,i} \}_{i \in \mathcal{K}_u}$ is the set of ordered UAV hovering locations to collect data from its associated CHs. 

The objective function is the total consumed energy by all UAVs, given their trajectories $\mathcal{L}_u$ associated to their sets of CHs $\mathcal{K}_u$. 
(\ref{c2}) ensures the successful communication between SNs in $\mathcal{A}_c$ and their associated CH $c$. 
(\ref{c1}) guarantees that enough energy is available in the battery to complete the mission at any TS $t$, while (\ref{c3}) satisfies the time deadline condition when a UAV collects data from CHs.
Also, (\ref{c4}) ensures that no collisions between UAVs occur, and finally, (\ref{c5}) limits the flying distance, for a given $v_{\max}$. 

Problem (P1) is NP-hard. Indeed, in the special case of already deployed CHs, the problem becomes finding the best routes for UAVs, while respecting the energy and time deadlines conditions.  The latter can be seen as the capacitated vehicle routing problem with time windows (CVRPTW) \cite{Lenstra1981}. The CVRPTW can be described as selecting the routes for a number of vehicles, aiming to serve a group of customers within time windows. Each vehicle has a limited capacity, which is used to depart from a depot point, serve a number of customers along its route, then return to the same depot point. The objective of the CVRPTW is to minimize the total transport costs. Logically, the vehicles, customers, and transport costs, can be assimilated by the UAVs, CHs, and energy consumption, respectively. Since the CVRPTW is known to be NP-hard \cite{Lenstra1981}, then by restriction, (P1) is NP-hard.

\section{Solution Approach}
In (P1), we notice that the optimization of parameters $\mathcal{U}$ and  $\{ \mathcal{K}_u, \mathcal{L}_u \}_{u \in \mathcal{U}}$ directly depends on the selected parameters $\mathcal{K}, \{ \mathcal{A}_c \}_{c \in \mathcal{K}}$ and $\mathcal{L}$. Moreover, since their associated constraints are independent, problem (P1) can be divided into two cascaded problems as follows:
\begin{enumerate}
    \item A first problem of sensors clustering and CHs placement, where only $\mathcal{K}$, $\mathcal{L}$, and $\{ \mathcal{A}_c \}_{c \in \mathcal{K}}$ are optimized, aiming to minimize the number of deployed CHs is formulated.\footnote{This objective has a direct impact on the consumed energy of data collection since a lower number of CHs would reduce the mission time and energy consumption of UAVs.}
    \item Then, given $\mathcal{K}$, $\mathcal{L}$ and $\{ \mathcal{A}_c \}_{c \in \mathcal{K}}$, a second problem of UAV trajectory planning is formulated in order to minimize the total consumed energy, through the optimization of $\mathcal{U}$ and $\{ \mathcal{K}_u, \mathcal{L}_u \}_{u \in \mathcal{U}}$.   
\end{enumerate}
Hence, we present next the two described problems and propose efficient approaches to solve them. 

 
\subsection{IoT Sensors Clustering}

Typically, IoT sensors are deployed in large areas in order to sense, process, and communicate relevant data for some applications. Due to their limited battery, IoT sensors need to carefully use their energy, while prolonging the life of collected data as much as possible. To do so, IoT sensors can be grouped into disjoint and non-overlapping clusters to reduce the amount of used energy. 
SNs scan the surrounding environment and transmit sensed data to CHs, which aggregate and transmit obtained information to the UAVs \cite{DBLP:conf/hicss/HeinzelmanCB00}.

Several state-of-the-art clustering techniques exist, e.g., K-means \cite{Kanungo2002}, density-based spatial clustering, and hierarchical cluster analysis (HCA) \cite{Nielsen2016}. Nevertheless, \textcolor{black}{due to the low-complexity of K-means,} we focus here on its customization, aiming to group SNs and deploy the minimal number of CHs. 
Thus, the associated clustering problem can be written as

\begin{subequations}
\small
	\begin{align}
	\label{p2}
	\min_{\substack{\mathcal{K},\mathcal{L},\\\{\mathcal{A}_c\}_{c \in \mathcal{K}}}} & \quad 
	\sum_{c=1}^{K} \sum_{s=1}^{M} a_{sc} \;  d_{sc}^2  \tag{P2} \\
	\label{c21} 
	\text{s.t.}\quad & d_{sc}\leq d_{\rm th}, \forall s \in \mathcal{A}_c, \forall c \in \mathcal{K}  \tag{P2.a} \\
	\label{c22} &  | \mathcal{A}_c| \leq F, \quad \forall c \in \mathcal{K},  \tag{P2.b}
	\end{align}
\end{subequations}
where $a_{sc} \in \{0,1\}$ is a binary variable indicating whether SN $s$ is associated with CH $c$ or not, $\forall s \in \mathcal{A}_c$. Constraint (\ref{c22}) guarantees fairness in associating SNs with CHs, where $F$ is the maximum number of SNs associated with one CH. Finally, $|\mathcal{A}_c|$ is the cardinality of the set $\mathcal{A}_c$.

Conventionally, K-means clustering requires a predefined number of CHs. Hence, in \cite{DBLP:conf/globecom/AlfattaniJYY19}, the authors proposed a K-means algorithm to solve (P2) without any constraint, then reexecuted it iteratively until (\ref{c21}) is met. Such method is time consuming and inaccurate, as the clustering performance may vary with the algorithm initialization. Therefore, we propose here to improve the method of \cite{DBLP:conf/globecom/AlfattaniJYY19} by integrating both constraints (\ref{c21})--(\ref{c22}) into the clustering process. The proposed approach is presented in Algorithm \ref{algo1}, and described as follows. First, locations of $K$ CHs are randomly initialized. Then, each SN is assigned to the closest CH with respect to (\ref{c21})--(\ref{c22}). Next, the locations of CHs are updated by calculating the resulting mean location of associated SNs for each CH. This procedure is repeated until convergence, i.e., the calculated locations remain unchanged. Since we aim to deploy the minimal number of CHs, we execute the aforementioned steps for an increasing number of CHs 
until a solution to (P2) is obtained, i.e., $K$ is determined.  

Although the calculated CHs locations are an adequate solution for (P2), this may not be the case for the main problem (P1). In this context, we propose to improve the locations of CHs by making them closer to the dockstation. Indeed, CH $c$ has a mobility margin if its furthest associated SN, denoted $s_0$, respects (\ref{c21}) loosely, i.e., $d_{s_0 c}< d_{\rm{th}}$. Hence, making the CHs closer to the dockstation would increase the probability to respect data collection deadlines, shorten the mission time, and improve the energy consumption of UAVs. This procedure is in lines 32-34 of Algorithm \ref{algo1}.


\begin{algorithm}[h]
\footnotesize
\caption{IoT sensors clustering algorithm}
\label{algo1}
\begin{algorithmic}[1]
\State \textbf{Input: }{$\mathcal{M}$, $\{\textbf{q}_s\}_{s \in \mathcal{M}}$}, and $\max_{\rm it}$ \; \%$\max_{\rm it}$ is the max. number of CHs
\State \textbf{Output: }{Optimal $\mathcal{K}$, $\mathcal{L}$, and $\{\mathcal{A}_c\}_{c \in \mathcal{K}}$}
\State Set {$k= ( |\mathcal{M}| \div F)$}
\For{$k$ \textbf{to} $\max_{\rm it}$} 
    \State Set $\mathcal{L}$ randomly \%Initial locations of CHs
    \State Set $\mathcal{L}_{old}$ \%Set of zeros
    \State Set $\mathcal{A}_c=\emptyset$, $\forall c=1,\ldots,k$  \%$c$ indicates the rank of \State CH in $\mathcal{L}$
    \State Set $error=dist(\mathcal{L},\mathcal{L}_{old})$ \; \%Convergence parameter 
    \While {$error \neq 0$ }
        \For {$s \in \mathcal{M}$}
            \State Calculate $d_{sc}$, $\forall c=1,\ldots,k$ 
            \State Find $c_0 = \arg (\min (d_{sc}))$
            \If {$d_{sc_0}\leq d_{\rm th} \And |\mathcal{A}_{c_0} \cup \{s\}|\leq F$}
                \State $\mathcal{A}_{c_0}=\mathcal{A}_{c_0} \cup \{s\}$ \%Associate SN $s$ with 
                \State CH $c_0$ 
            \Else 
                \State $\mathcal{A}_{k+1}=\mathcal{A}_{k+1} \cup \{s\}$ \%Put $c$ in set of no 
                \State association 
                \EndIf
            \EndFor
        \State Set $\mathcal{L}_{old}=\mathcal{L}$
        \State Calculate $\mathcal{L}$ as the mean location of the associated 
        \State SNs in $\mathcal{A}_c$, $\forall c=1,\ldots,k$            
        \State Calculate $error=dist(\mathcal{L},\mathcal{L}_{old})$  
    \EndWhile
    \If{$\mathcal{A}_{k+1} =\emptyset$}
        \State $\mathcal{K}=\{1,\ldots,k\}$
        \State Break
    \EndIf
\EndFor
\While{$\max(d_{sc}) < d_{\rm th}, \; \forall s \in \mathcal{A}_c$}
    \State Get $\mathcal{L}$ closer to the dockstation
\EndWhile
\end{algorithmic}
\end{algorithm}

 
\subsection{Multi-UAV Trajectory Planning}
Given $\mathcal{K}$, $\mathcal{L}$, and $\mathcal{A}_c$, $\forall c \in \mathcal{K}$, the trajectory planning problem can be formulated as 

\begin{subequations}
\small
\begin{align}
\label{p3} 
\min_{\substack{\mathcal{U},\{\mathcal{L}_u,\mathcal{K}_u}\}_{u \in \mathcal{U}}} &  \quad 
\sum_{u=1}^U E_u(\mathcal{K}_u) \tag{P3} \\
\text{s.t.}\quad & (\text{\ref{c1}})- (\text{\ref{c5}}). \notag
\end{align}
\end{subequations}
As discussed previously, this problem is NP-hard, and can be assimilated to a CVRPTW problem.
In order to solve it, we model our system as a graph, described as follows. Let $G = (\mathcal{D},\mathcal{E})$ be a complete graph, where $\mathcal{D} = \{0, 1, \ldots, K\}$ is a set of vertices (nodes) representing the dockstation (node 0) and $K$ CHs, and $\mathcal{E}$ is the set of directed edges connecting the nodes. A directed edge from node $i$ to node $j$, denoted $e_{ij}$, represents the UAV's flying operation from $i$ to $j$ and the hovering operation at $j$. A cost associated to the edge $e_{ij}$, called $\chi_u(e_{ij})=E_u(v_u,\mu_1(e_{ij}))+E_u(0,\mu_2(e_{ij}))$, is expressed as the sum of the UAV's flying and hovering energy, where $\mu_1(e_{ij})=T_{f,u}(i,j) \delta$ and $\mu_2(e_{ij})=T_{ju} \delta$ respectively, $\forall u \in \mathcal{U}$.\footnote{This is valid assuming that all UAVs are of the same type, i.e., having the same mechanical and communication characteristics.} Moreover, the time deadlines at CHs are represented in the graph by a time window $\omega_c=[o_c, v_c]$, where $o_c$ and $v_c$ are the minimum and maximum instants for data collection, for each node $c \in \mathcal{D}$. This time window defines when data collection at node $c$ can begin and end. Finally, trajectory of UAV $u$ in $\mathcal{G}$ can be defined by $\mathcal{K}_u$, where each element of $\mathcal{K}_u$ is the ordered node to visit.    
Accordingly, (P3) can be reformulated as problem (P4) detailed below: 

\begin{subequations}
\small
	\begin{align}
	\label{p4}
	\min_{\substack{\mathcal{U}, \{\mathcal{K}_u, \\  \mathcal{L}_u\}_{u \in \mathcal{U}}}} & \quad 
	\sum_{u\in \mathcal{U}} \sum_{i \in \mathcal{D}} \sum_{j \in \mathcal{D}} 
	\chi_u(e_{ij})\; b_{u,ij}  \tag{P4} \\
	\label{c41} 
	\text{s.t.}\quad & \sum_{u=1}^{U} \sum_{i \in \mathcal{D}} b_{u,ij} \leq 1, \quad \forall j \in \mathcal{D}  \tag{P4.a} \\
	\label{c42} & \sum_{j \in \mathcal{D}} b_{u,0j} = \sum_{i \in \mathcal{D}} b_{u,i0} = 1, \quad \forall u \in \mathcal{U}  \tag{P4.b} \\
	\label{c43} & \sum_{i=1}^{j-1} \mu_1(e_{i(i+1)})+\mu_2(e_{i(i+1)}) \in [o_c,v_c], \; \forall j \in \mathcal{K}_u \tag{P4.c} \\
	\label{c44} & \sum_{i \in \mathcal{D}} \sum_{j \in \mathcal{D}} 
	\chi_u(e_{ij}) b_{u,ij} \leq S_u(0)-S_{\min}, \; \forall u \in \mathcal{U}, \tag{P4.d}
	\end{align}
\end{subequations}
where $b_{u,ij}$ is the binary variable indicating the selection of the edge $i-j$ within UAV $u$'s trajectory.
Constraint (\ref{c41}) ensures that each CH is visited at most once by exactly one UAV. (\ref{c42}) guarantees that each UAV departs and returns to the dockstation. Condition (\ref{c43}) emphasizes that the data collection time cannot exceed the time deadline. Finally, (\ref{c44}) guarantees that the consumed energy by any UAV respects the battery capacity.
Subsequently, problem (P4) can be solved using any of the available CVRPTW heuristic or metaheuristic approaches, such as gradient descent, simulated annealing, Tabu search, etc. \cite{GoogleOR}.

\section{Numerical Results}

We assume $M$=2000 IoT sensors randomly and uniformly distributed within a geographical area of $5 \times 5$ km$^2$. 
Unless specified otherwise, 
we use the simulation parameters of
Table \ref{tab:table1}, as presented in the next page.



\begin{figure*}
\begin{minipage}{0.315\linewidth}
{\includegraphics[width=144pt]{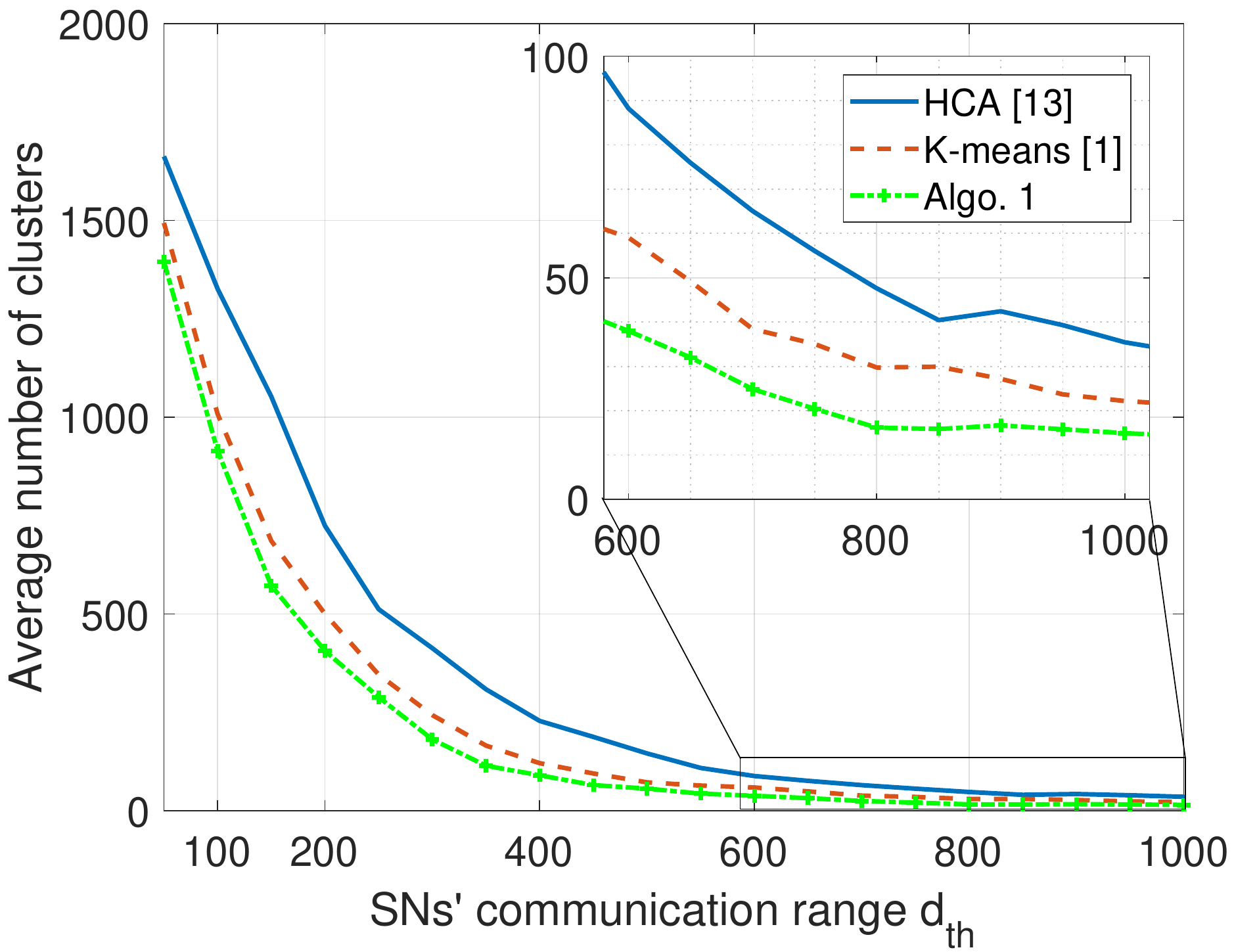}}
\caption{Avg. no. of clusters vs. $d_{\rm th}$.}
\label{Fig1}
\vfill
{\includegraphics[width=144pt]{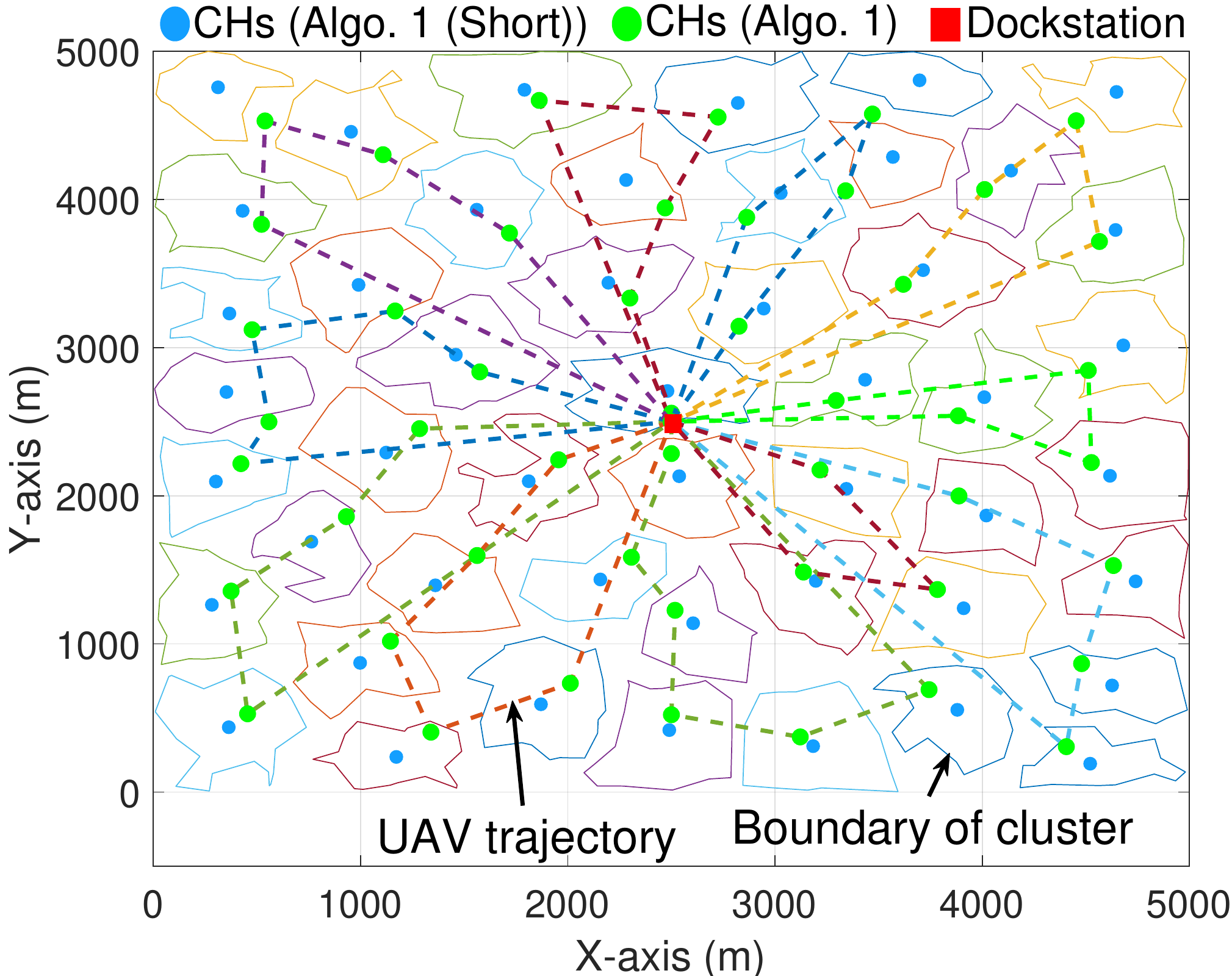}}
\caption{Clustering and UAV trajectories.}
\label{Fig2}
\end{minipage}%
\hfill
\begin{minipage}{0.315\linewidth}
{\includegraphics[width=144pt]{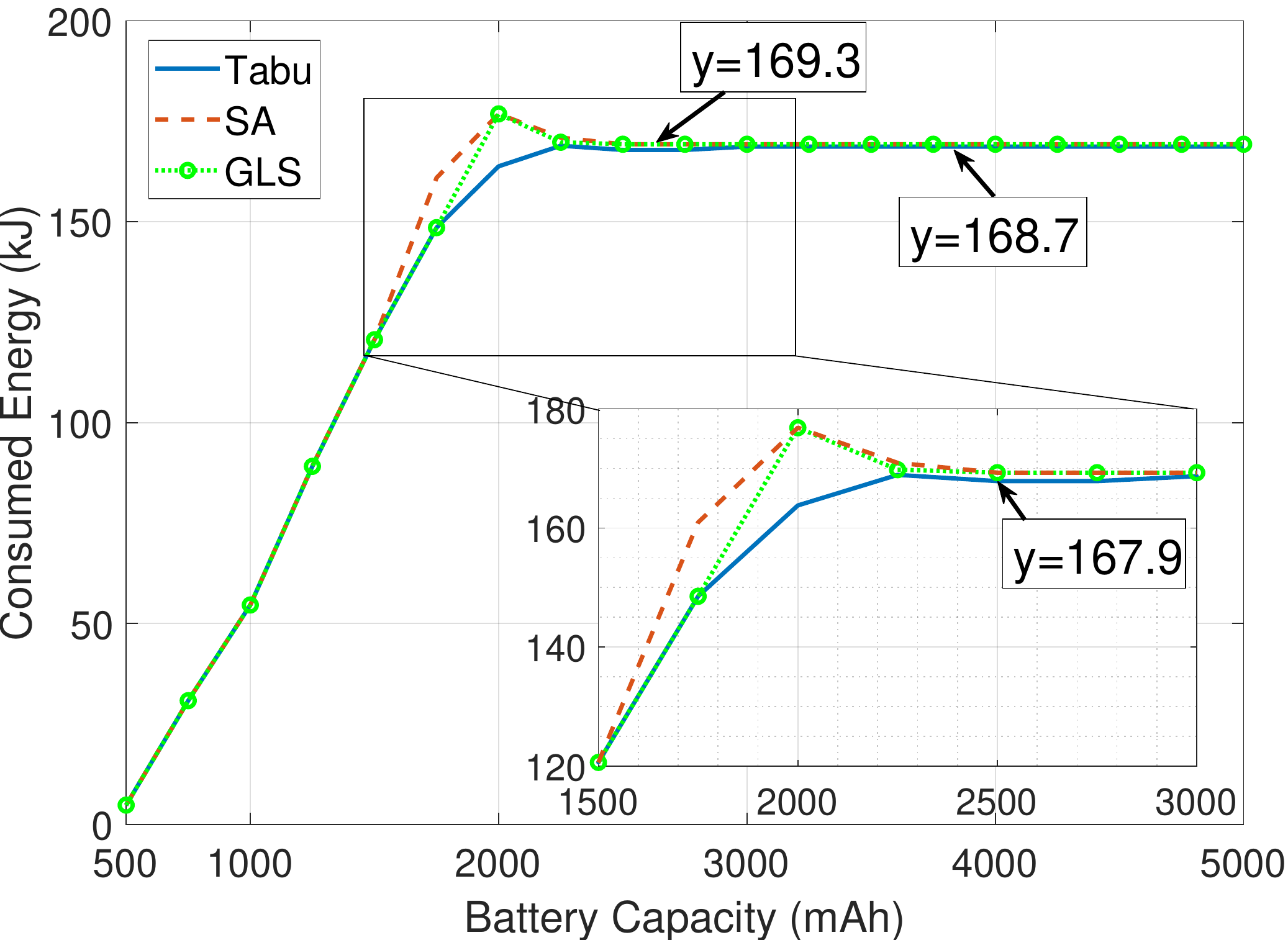}}
\caption{Energy vs. battery capacity.}
\label{Fig3}
\vfill
{\includegraphics[width=144pt]{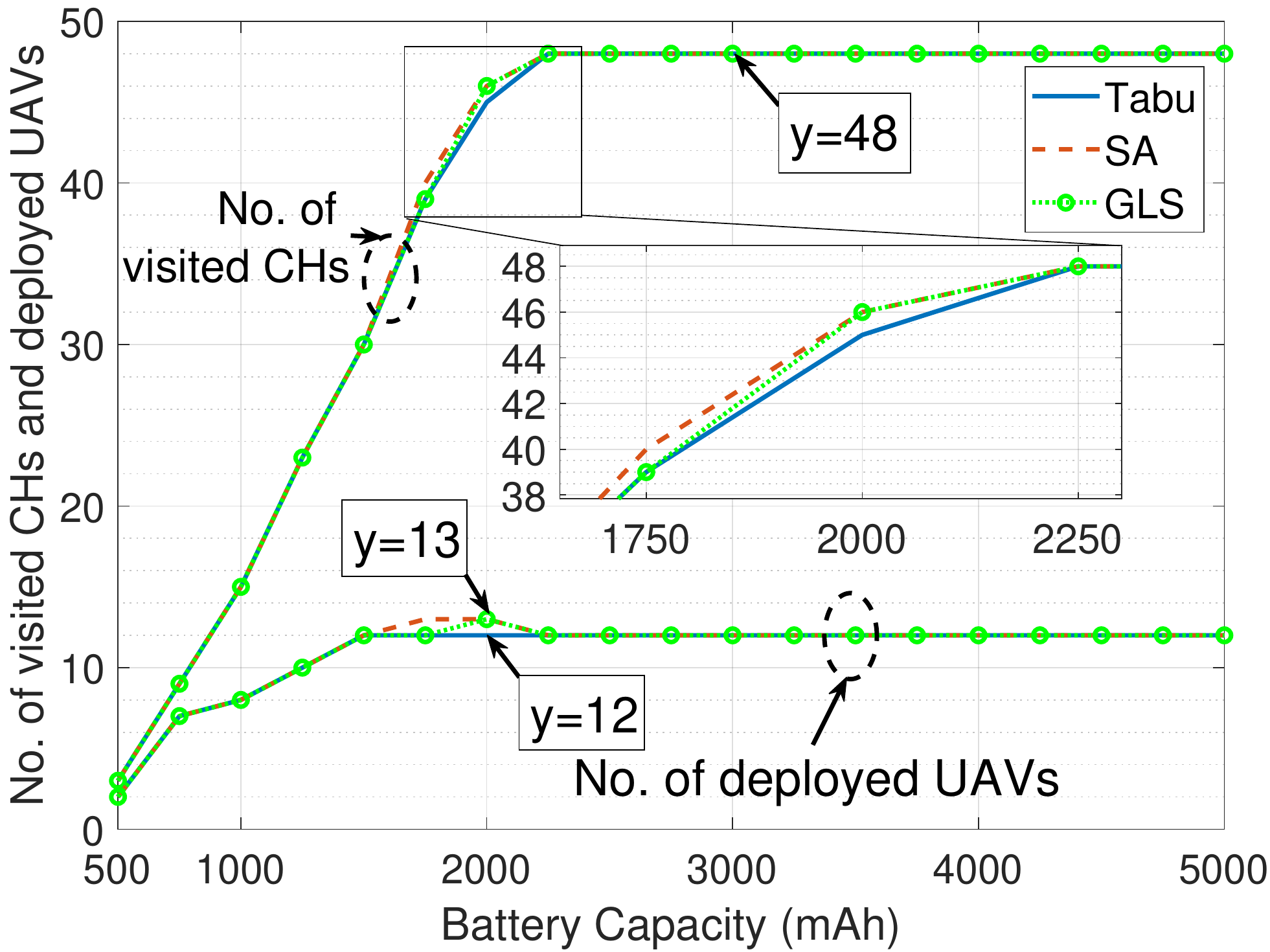}}
\caption{No. of CHs and UAVs vs. battery capacity.}
\label{Fig4}
\end{minipage}%
\hfill
\begin{minipage}{0.315\linewidth}
{\includegraphics[width=144pt]{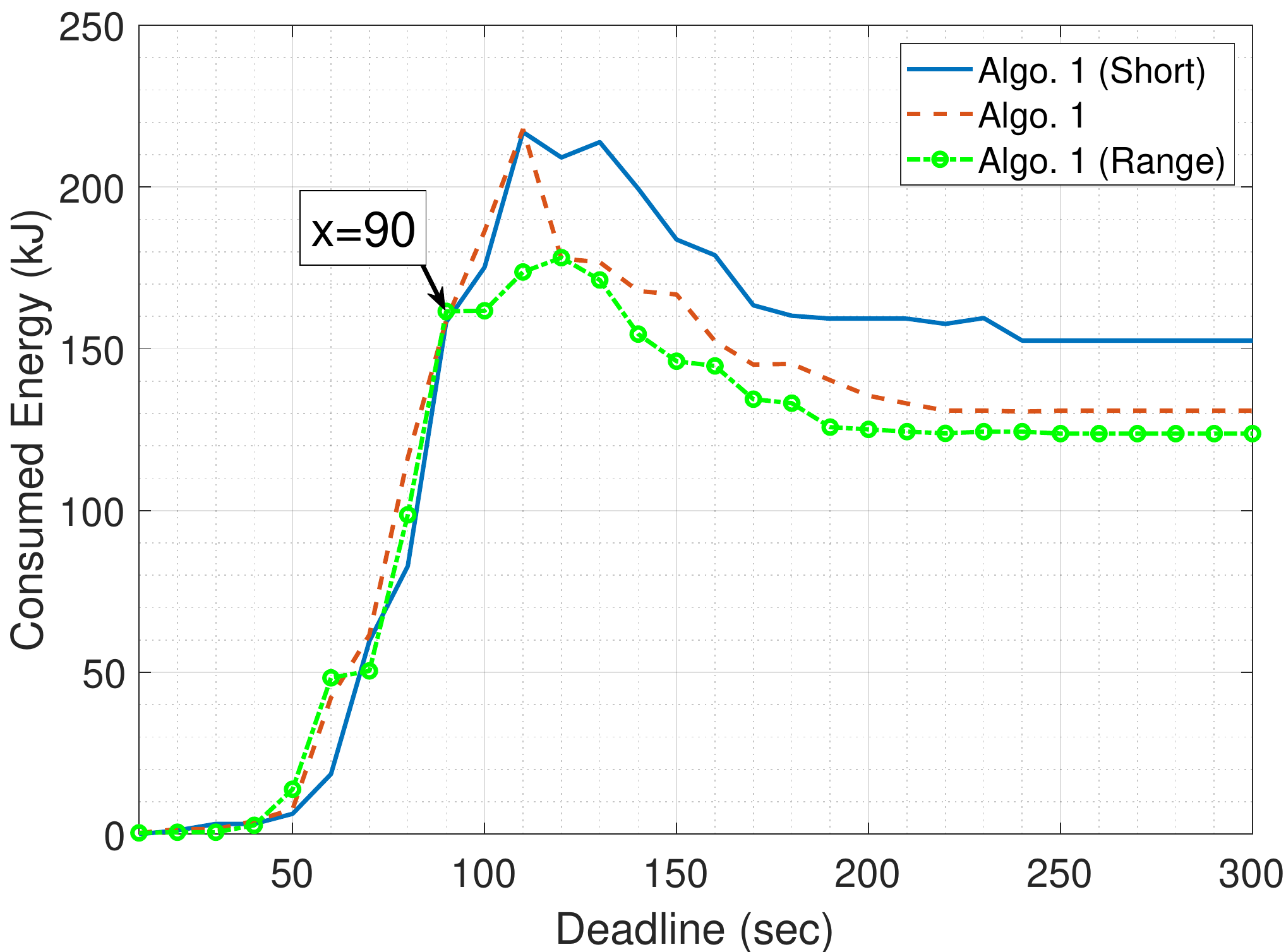}}
\caption{Energy vs. deadline.}
\label{Fig5}
\vfill
{\includegraphics[width=144pt]{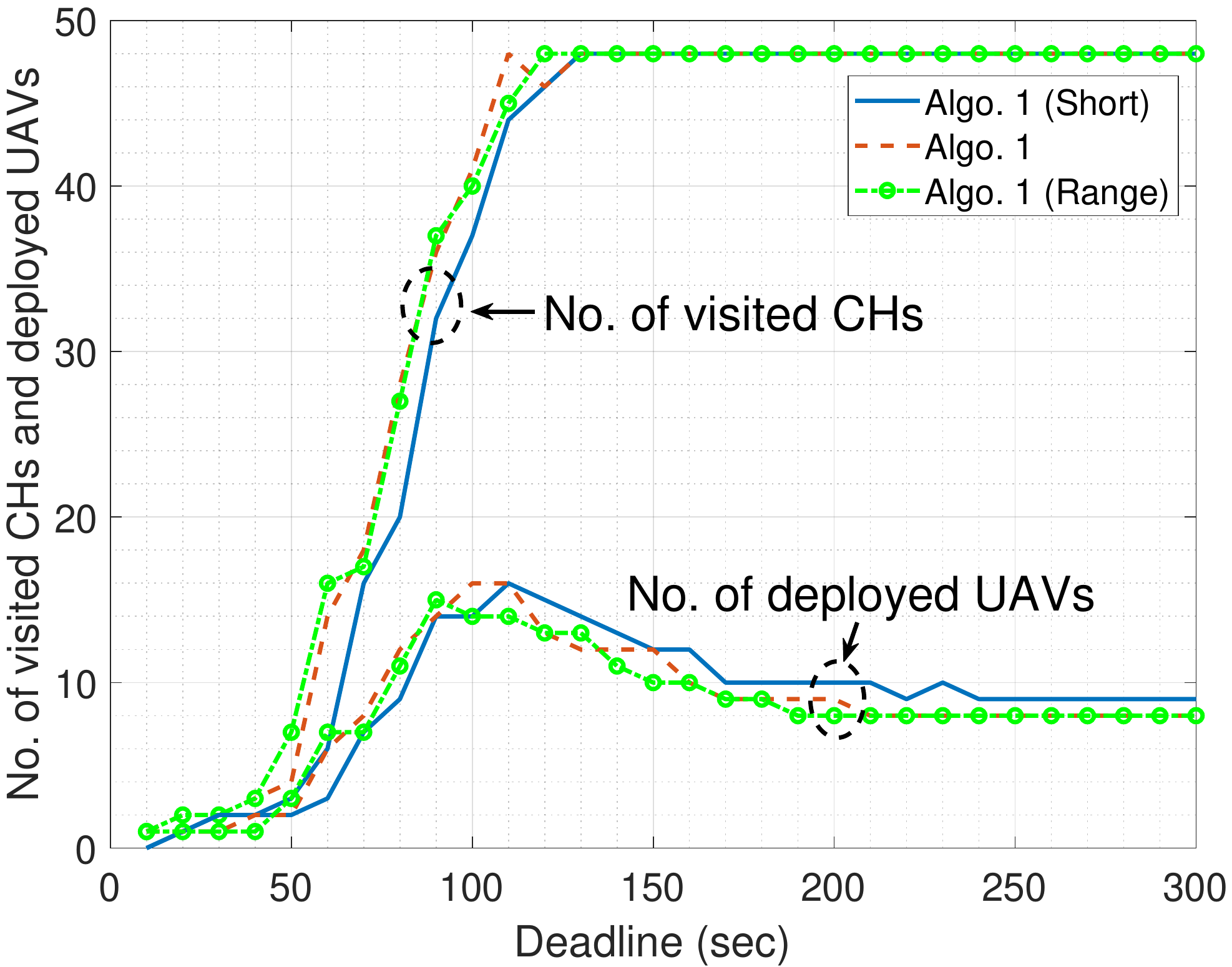}}
\caption{No. of CHs and UAVs vs. deadline.}
\label{Fig6}
\end{minipage}%
\end{figure*}

Fig. \ref{Fig1} presents the clustering performance of several algorithms, expressed by the average number of clusters as a function of communication range, $d_{\rm th}$ (averaging over 100 scenarios of SNs). We consider three clustering algorithms, namely proposed Algo. 1, K-means-based clustering in \cite{DBLP:conf/globecom/AlfattaniJYY19} (K-means Algo. \cite{DBLP:conf/globecom/AlfattaniJYY19}), and HCA-based clustering  (HCA \cite{Nielsen2016}). As $d_{\rm th}$ increases, a smaller number of CHs is needed, as the latter can be reached by a higher number of SNs. Moreover, Algo. 1 achieves the best clustering performance. Indeed, unlike the other algorithms, Algo. 1 is capable of adjusting the clusters to constraints (\ref{c21})--(\ref{c22}) on-the-fly, i.e., while processing the SNs.     

In Fig. \ref{Fig2}, we depict a scenario where Algo. 1 is used to group 2000 SNs into 48 clusters, from which data is collected using 11 UAVs. We distinguish two variations, Algo. 1 as is, and ``Algo. 1 (Short)", where lines 32--34 are omitted. It is obvious that Algo. 1 would result in a more energy-efficient data collection, since UAVs would fly for less time, assuming that they hover directly above each CH to collect data.   

For the clustering design of Fig. \ref{Fig2}, Figs. \ref{Fig3}--\ref{Fig4} compare the consumed energy (in kilojoule), number of visited CHs, and number of deployed UAVs performances, as functions of the battery capacity, for different UAV trajectory planning approaches, namely Tabu search (Tabu), simulated annealing (SA), and guided local search (GLS). Performances of these methods are almost similar, with a preference for Tabu. Indeed, Tabu consumes less or equal energy to SA and GLS. \textcolor{black}{Also, as the battery capacity increases (above 2250 mAh), the performance saturates, guaranteeing that data is collected from all CHs. Finally, the optimal battery capacity of 2500 mAh is provided by Tabu, where minimum energy (=167.9 kJ) and number of UAVs (=12) are achieved.}

In Figs. \ref{Fig5}--\ref{Fig6}, the same aforementioned performances are depicted as functions of the time deadline, for variations of Algo. 1, including ``Algo. 1 (Short)" and ``Algo. 1 (Range)", where in the latter, UAVs do not hover directly above CHs, but rather within a range {of 150 m from the initial UAV locations defined in Algo. 1}. 
For all approaches, as the deadline increases, the consumed energy reaches a peak then decreases. Below this peak, the deadline is very small such that it prevents collecting data from all CHs as the latter will be outdated, hence, a small number of UAVs is deployed, which consumes low energy due to short trajectories. In contrast, beyond the peak point, the deadline is long enough to deploy few UAVs that visit all CHs. The peak point corresponds to the critical deadline for which the maximum number of UAVs is deployed to collect data from the CHs. ``Algo. 1 (Range)" presents the best energy consumption for deadlines above 90 sec, while visiting and deploying similar numbers of CHs and UAVs respectively, as Algo. 1. Indeed, ``Algo. 1 (Range)" compensates for the prolonged data collecting time and energy (due to longer distances to CHs) by shorter flying time and energy. For deadlines below 90 sec, Algo. 1 performs either better or similarly to ``Algo. 1 (Range)", in terms of energy and number of visited CHs, since hovering exactly above CHs improves the data transmission and thus respects the deadline. Finally, ``Algo. 1 (Short)" presents the worst performances as it spends more energy to reach CHs at distant locations, or in contrast, abandon them due to their data becoming outdated.  
\begin{table}[ht]
\scriptsize
\caption{Simulation parameters }
\centering
\renewcommand{\arraystretch}{1}
\begin{tabular}{|c|c||c|c|} 
\hline
UAV altitude& $H$=100 m  & Bandwidth & $W$=10 MHz \\ 
\hline 
\makecell{UAV speed}& \makecell{$v_{u}$=30 m/sec} & \makecell{Speed \\of light} & \makecell{$V$=3 $\times10^8$ m/sec} \\
\hline
\makecell{Distance \\threshold} & \makecell{$d_{\rm th}$=600 m}  & \makecell{Mission \\ time}& \makecell{$T_f$=4 min} \\ 
\hline
\makecell{Carrier \\frequency}& \makecell{$f_c$=2 GHz} & \makecell{Environment\\ parameters}& \makecell{$a$=9.61\\ $b$=0.16} \\ \hline
\makecell{Time\\deadline} & \makecell{$T_{d,c}$=140 sec} & \makecell{Max SNs\\per cluster} & \makecell{$F$=120}\\
\hline
Noise power& $\sigma^2$=--109 dBm & Packet size & $S_p$=1 Kbyte  \\ 
\hline
\makecell{TS length}& \makecell{$\delta$=100 msec} & \makecell{Path loss} & \makecell{$\alpha$=2.7}\\
\hline
\makecell{Battery \\capacity}& \makecell{$S_u(0)$=3,500 mAh} & \makecell{Number \\of packets}& \makecell{$Q_c$=50,000}\\
\hline
\end{tabular}
\label{tab:table1}
\end{table}

\section{Conclusion}
In this paper, we investigated the multi-UAV data collection problem in clustered IoT networks, where sensed data have time deadlines. Aiming to optimize the data collection deployment costs in terms of energy consumption, we propose a two-step solution. In the first step, the number and locations of CHs, which gather data from associated IoT sensors, are optimized using a customized K-means approach. 
Subsequently, an energy-efficient data collection framework which uses the minimal number of UAVs is presented; in this framework, the UAV trajectories are defined with respect to time deadlines of 
collected data and energy constraints.
Simulation results show the efficiency of our customized K-means clustering compared to baseline approaches. Also, moving CHs closer to the dockstation provided a significant energy gain. On the other hand, it is shown that Tabu search achieves the best UAV trajectory design, compared to other methods. Finally, the impact of the battery capacity and time deadline is studied in terms of energy consumption, number of visited CHs, and number of deployed UAVs.        

\bibliographystyle{IEEEtran}
\bibliography{IEEEabrv,tau}

\end{document}